# Electric Field and SAR Reduction in High Impedance RF Arrays by Using High Permittivity Materials for 7T MR Imaging

Aditya A Bhosale[1], Yunkun Zhao[1], Xiaoliang Zhang[1,2*]

[1]Department of Biomedical Engineering, State University of New York at Buffalo, Buffalo, NY, United States
[2]Department of Electrical Engineering, State University of New York at Buffalo, Buffalo, NY, United States
[*]Corresponding Author: xzhang89@buffalo.edu

*Abstract*— **Higher frequencies and shorter wavelengths present significant design issues at ultra-high fields, making multi-channel array setup a critical component for ultra-high field MR imaging. The requirement for multi-channel arrays, as well as ongoing efforts to increase the number of channels in an array, are always limited by the major issue known as inter-element coupling. This coupling affects the current and field distribution, noise correlation between channels, and frequency of array elements, lowering imaging quality and performance. To realize the full potential of UHF MRI, we must ensure that the coupling between array elements is kept to a minimum. High-impedance coils allow array systems to completely realize their potential by providing optimal isolation while requiring minimal design modifications. These minor design changes, which demand the use of low capacitance on the conventional loop to induce elevated impedance, result in a significant safety hazard that cannot be overlooked. High electric fields are formed across these low capacitance lumped elements, which may result in higher SAR values in the imaging subject, depositing more power and, ultimately, providing a greater risk of tissue heating-related injury to the human sample. We propose an innovative method of utilizing high-dielectric material to effectively reduce electric fields and SAR values in the imaging sample while preserving the B1 efficiency and inter-element decoupling between the array elements to address this important safety concern with minimal changes to the existing array design comprising high-impedance coils.**

## I. INTRODUCTION

Multi-channel arrays outweigh volume coils in terms of improved signal-to-noise ratio and faster acquisition when used in combination with parallel imaging techniques [1-7]. At ultra-high-field magnetic resonance imaging (7T and above), the Larmor frequency increases and the wavelength decreases, making it challenging to develop larger imaging coils, such as volume coils [8-12]. The addition of the channel count in a multi-channel array is regarded as advantageous due to the positive impact of higher channel numbers on the signal-to-noise ratio (SNR), consequently leading to an enhancement in the magnetic resonance (MR) image, but more channels also present a technical challenge known as inter-element coupling [13-18]. Inter-element coupling impacts the current and B1 field distribution and contributes to correlated noise that degrades SNR and parallel imaging performance [19-25]. It is critical to keep inter-element coupling to a minimum to preserve imaging performance. Numerous efforts have been made to reduce inter-element coupling between array elements while maintaining the present element count. One well-known example is overlapping neighboring elements in arrays to minimize interaction between the surrounding components [26]. In receive-only arrays, this method is generally used together with a low input impedance pre-amplifier decoupling method to reduce interaction between non-adjacent elements although it is not readily feasible for transceiver arrays [13, 26]. Capacitive/inductive decoupling networks constitute additional methods for improving inter-element isolation between array elements [27-29]. Other attempts have been made to reduce coupling by utilizing the metamaterial substrate to reduce or eliminate the induced currents [30-41]. All of the aforementioned approaches provide excellent inter-element isolation, which improves imaging performance but also increases the complexity of array designs due to the added decoupling circuitry. The high-impedance coil design is another decoupling solution that has been revisited and further investigated recently [26, 42-46]. This method not only improves inter-element isolation but also simplifies array construction by eliminating the need for additional circuitry, leading to a robust array design method with high efficiency and durability, particularly for flexible multichannel arrays. To introduce high impedance into the resonator circuit, small values of lumped capacitors are often selected. High electric fields are created across these low-capacitance capacitors. Because the specific absorption rate (SAR) is proportional to the square of the electric fields, higher electric fields imply higher SAR values [47, 48]. SAR is a measure of how much power an RF field deposits in a certain mass of tissue and is a key cause of tissue heating in MR imaging. As a result, keeping SAR values within a specific range is crucial to ensuring human safety during MR imaging tests. High-impedance coils preserve the geometry of the conventional loop while offering excellent inter-element isolation between array elements. However, the high impedance property may result in high electric field generation over the lumped elements with low capacitance, resulting in high E-spots on the coil and higher power being deposited into the tissue, causing tissue heating or burns, a known safety hazard that must be addressed properly. To address this issue, we propose to utilize the known potential of the high dielectric constant material [49-51]. High-dielectric-constant materials have been shown to absorb E-fields and positively alter B1 field distributions [52-59]. In this work, we develop and investigate a method of using a thin, high-dielectric sheet to minimize electric fields and SAR values while preserving inter-element isolation and B1 efficiency of the high-impedance coil arrays. Numerical simulation analysis was employed to assess scattering parameters, electric fields, SAR, and B1 fields across different human tissue properties in experimental scenarios. These scenarios involved varying sizes and relative permittivity values of a high dielectric constant material sheet, as well as its distance from high-impedance coils. Additional validation of the methodology was conducted through the construction of a prototype and the subsequent execution of bench tests in order to demonstrate the viability of the concept.

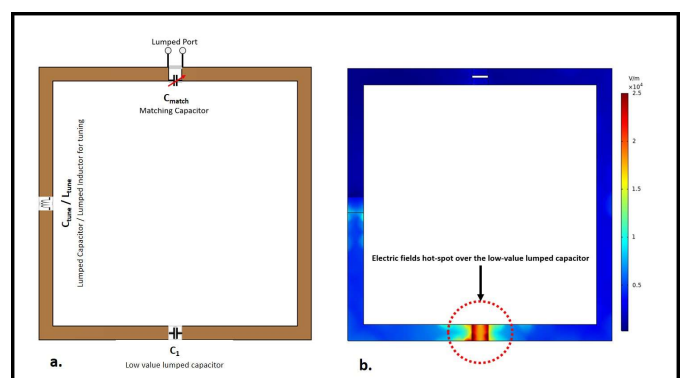

Figure 1. (a) Circuit schematic of a high-impedance coil (b) Surface electric field distribution over the single high-impedance coil element. Higher electric fields are observed over the low-value lumped capacitor placed opposite the port.

## II. METHODS

This section goes over the several cases we investigated for our proposed technique. A high-impedance coil without any high dielectric constant material is looked at and compared to an instance with high dielectric constant material placed between the phantom and the high-impedance coil. We chose two dimensions for the high dielectric constant material, which would be expanded upon in the study. To verify that the coil was tuned at 300 MHz and precisely matched at 50 ohms, the relative permittivity, and distance of the high dielectric constant material from the coil were varied within a specific range. A cylindrical phantom with a 30 cm diameter and 30 cm height was used to evaluate the inter-element isolation, electric field, B1+ fields, and SAR values for each case. The phantom was assigned different tissue parameters, such as the human brain, breast fat, kidney, and tendon/ligament, to evaluate all the resultant parameters in depth.

### A. High-impedance coils without the high dielectric constant material

The inter-element isolation of two $10 \times 10 cm^2$ high-impedance coil resonators was evaluated. The 1 cm distance between the two resonators was maintained. A cylindrical phantom with a diameter of 30 cm and a height of 30 cm was used and assigned the properties of human brain tissue (conductivity $\sigma$ = 0.6 S/m and permittivity $\varepsilon_r$ = 50). The same 1.5 cm distance was maintained between the cylindrical phantom and the resonators. By selecting the appropriate capacitors and inductors, the high-impedance coils were tuned to 300 MHz, and their impedances were matched to 50 ohms. The elements had a low-value Cmode capacitor (0.35 pF) placed opposite the feed port, which allowed the resonators to attain a high impedance and create a dipole-like open-path current pattern for exceptional decoupling behavior. Schematic co-simulation from Dassault Systemes' CST studio suite was used to precisely select the low-value Cmode capacitor and Xarm lumped inductor (40 nH) to tune the coil at 300 MHz and achieve the self-decoupling property of the high-impedance coil. In addition, a shunt Cmatch capacitor (15 pF) was connected to the input port to match the impedance of each channel to 50 ohms. **Fig. 1.a** depicts the circuit diagram of the high-impedance coil and the positioning of each lumped element on the coil. The low capacitance required for decoupling results in an increased E-field being generated across it. The surface electric field distribution over the high-impedance coil element is depicted in **Fig. 1.b.** Increased electric fields are observed close to the low-value lumped capacitor, as depicted in the picture. To address this issue, we tested numerous scenarios involving the use of high-dielectric-constant materials. The complete simulation setup used to evaluate the high-impedance coils without the high dielectric constant material can be seen in **Fig. 2.**

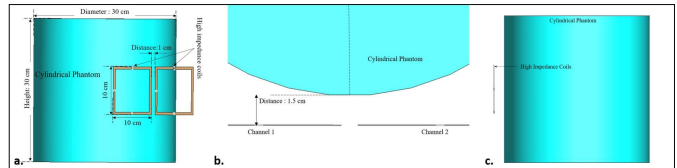

Figure 2. Simulation setup for the high-impedance coil without the high dielectric constant material (a) A 3-dimensional view of the setup showing the two high-impedance resonators placed 1.5 cm away from the cylindrical phantom (b) The top view of the setup shows the position of each component used in the simulation setup. (c) A side view of the simulation setup.

### B. $30 \times 11.5 cm^2$ High dielectric constant material sheet partially covering the high impedance coils

Minor changes were made to the previous design by introducing a $30 \times 11.5 cm^2$ high dielectric constant (HDC) material sheet with a thickness of 1 mm between the cylindrical phantom and the high-impedance coils. **Fig. 3** depicts the complete simulation setup for the case. Phantom's material properties remained consistent with the previous design. The distance between the phantom and the high-impedance coils was kept at 1.5 cm, as in the previous case. The HDC material was positioned between these two components, with its distance from the high-impedance coils varying from 3 mm to 8 mm. The relative permittivity of the HDC material was modified, and several values ranging from 50 to 200 were utilized to evaluate the material's impacts on inter-element isolation and other factors. When the HDC material is placed in front of the high-impedance coil, it introduces a dielectric load to the resonator, causing the tuning frequency to shift by a small margin. As a result, adjusting the impedances responsible for matching the tuning frequency for each situation suitably is critical. The high-impedance coils were easily fine-tuned at 300 MHz by changing the Xarm impedances, but it was also clear that the low-value Cmode capacitor needed to be changed to find its appropriate combination with the Xarm inductors to enable the high-impedance coil's self-decoupling property. The shortest distance between the HDC material and the resonator tested was 3 mm, and the others were 5 mm, 8 mm, and 10 mm. The distances were varied to compare the reductions in electric fields based on the distance from the HDC material. The relative permittivity values tested for each distance varied because the dielectric load imposed on the resonator was determined by the proximity of the HDC material, making testing the higher permittivity values in the closest distance cases difficult.

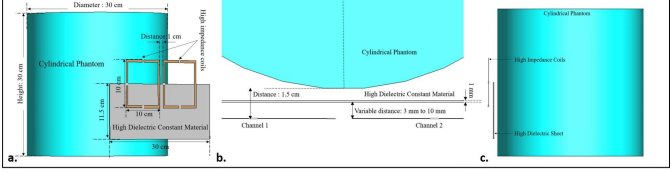

Figure 3. Simulation setup for the high-impedance coils with the 30 × 11.5 cm² HDC material sheet partially covering them (a) A 3-dimensional view of the setup showing the two high-impedance resonators placed 1.5 cm away from the cylindrical phantom and the 1 mm thick HDC material sheet placed between them (b) A top view of the setup showing the position of each component used in the simulation (c) A side view of the simulation setup

| Distance | relative permittivity ($\varepsilon_r$) | Cmode (pF) | Xarm (nH) | Cmatch (pF) |
|---|---|---|---|---|
| 3 mm | 50 | 0.2355 pF | 15.5 nH | 16.3 pF |
|  | 100 | 0.1 pF | 22 nH | 15.4 pF |
| 5 mm | 50 | 0.24 pF | 32 nH | 15.4 pF |
|  | 100 | 0.11 pF | 45 nH | 15 pF |
| 8 mm | 50 | 0.2 pF | 44 nH | 15.7 pF |
|  | 100 | 0.15 pF | 54.6 nH | 15.1 pF |
|  | 150 | 0.09 pF | 62.7 nH | 14.8 pF |
| 10 mm | 50 | 0.265 pF | 42 nH | 15.7 pF |
|  | 100 | 0.195 pF | 52 nH | 15.1 pF |
|  | 150 | 0.125 pF | 63 nH | 15 pF |
|  | 200 | 0.09 pF | 68 nH | 14.7 pF |

Table 1. Lumped element values used for each case of 30 × 11.5 cm² HDC material used over the high-impedance coils

The experiment attempted to reduce the higher electric field generation by adding a thin layer of HDC material over the coil at a specific distance while keeping the geometry and placement of each lumped component on the coil unchanged. The values of the lumped components used to tune the high-impedance resonators tuned at 300 MHz and matched at 50 ohms are listed in **Table. 1.** Schematic circuit co-simulation was used to identify the combination of Cmode and Xarm impedances at which the self-decoupling behavior of the resonators was observed.

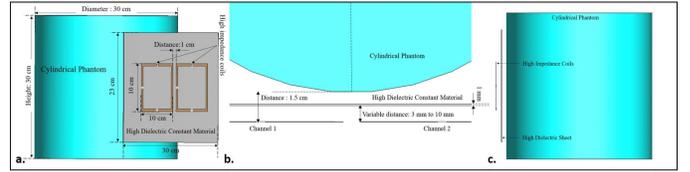

Figure 4. Simulation setup for the high-impedance coils with the 30 × 23 cm² HDC material sheet partially covering them (a) A 3-dimensional view of the setup showing the two high-impedance resonators placed 1.5 cm away from the cylindrical phantom and the 1 mm thick HDC material sheet placed between them (b) A top view of the setup showing the position of each component used in the simulation (c) A side view of the simulation setup.

### C. 30 × 23 cm² High dielectric constant material sheet completely covering the high-impedance coils

The dimensions of the high dielectric constant material sheet used in the previous design were increased to 30 × 23 cm² while keeping the thickness the same at 1 mm. The rest of the simulation setup, including the cylindrical phantom and the high-impedance coils, remains consistent with previous cases. Similar to the prior case, the following distances between the HDC material and the high-impedance resonators were tested: 3 mm, 5 mm, 8 mm, and 10 mm, and the relative permittivity values at which the resonators tune at 300 MHz and preserve the decoupling performance were tested depending on the proximity of the HDC material to the resonators. **Fig.4.** shows the simulation setup used for the evaluation of the 30 × 23 cm² HDC material covering the high-impedance coils completely while placed at a certain distance from the resonators.

Following the same approach as the previous instance, each case evaluation required tuning the high-impedance resonators at 300 MHz and impedance matching at 50 ohms. **Table.2.** shows the values that were employed in combination to achieve frequency tuning and self-decoupling behavior.

| Distance | Relative permittivity ($\varepsilon_r$) | Cmode (pF) | Xarm (nH) | Cmatch (pF) |
|---|---|---|---|---|
| 3 mm | 50 | 0.265 pF | 2 nH | 17.4 pF |
|  | 100 | 0.15 pF | 4 nH | 16.1 pF |
| 5 mm | 50 | 0.245 pF | 25 nH | 17 pF |
|  | 100 | 0.145 pF | 33 nH | 15.5 pF |
| 8 mm | 50 | 0.265 pF | 35 nH | 16.7 pF |
|  | 100 | 0.18 pF | 45 nH | 15.4 pF |
|  | 150 | 0.1 pF | 57 nH | 14.7 pF |
| 10 mm | 50 | 0.265 pF | 40.8 nH | 15.5 pF |
|  | 100 | 0.195 pF | 50.5 nH | 15.2 pF |
|  | 150 | 0.15 pF | 54.5 nH | 14.5 pF |
|  | 200 | 0.11 pF | 59.28 nH | 14.1 pF |

Table 2. Lumped element values used for each case of 30 × 23 cm² HDC material used over the high-impedance coils

### D. Construction and bench test measurements

Using a 3D printer, a 3D model of the 10 × 10 cm² high-impedance coil was created. To replicate the coil design, the copper tape was employed as the conductor and adhered to the printed PLA former. In addition, the Cmode was a low-value trimmer capacitor (Johanson Giga-trim JK-272 0.4-2.5pF) that was mounted opposite the coaxial line feed. For tuning the coil at 300 MHz, a 10-pF variable capacitor was employed as an Xarm impedance. A shunt capacitor was also added to the feed line to match the coil's impedance at 50 ohms, much like in the numerical simulation setup. We created a 3D-printed tray with the dimensions 28 × 22 cm$^2$ and 28 × 11 cm$^2$ to cover the high-impedance coil models to recreate the simulation scenario. Each of the printed trays was then covered with a 1 mm coating of gelatin and distilled water solution to make an HDC material with approximate relative permittivity ($\varepsilon_r$) of 78. The bench testing arrangement and the appropriate simulation setup for every case investigated are shown in **Fig. 5**. The H-field and E-field probes were connected to a 3D positioning system that was constructed using a high-precision CNC router (Genmitsu PROVerXL 4030) to map the corresponding fields above the high-impedance coil. The field mappings were evaluated based on the transmission coefficient S-parameters acquired through the probes at different points in a specified plane above the coil. The raw data transmission and the reflection coefficient values were obtained using the vector network analyzer (Keysight, E5061B, Santa Clara, CA, USA) and the data were further processed using MATLAB to acquire the field maps.

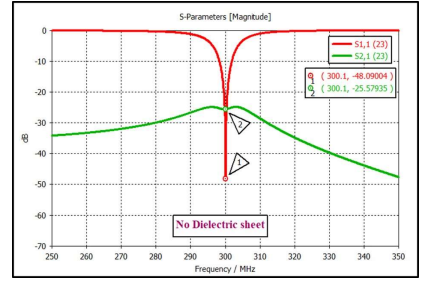

Figure 6. Scattering parameters show the reflection and the transmission coefficient of the high-impedance coils without the HDC material.

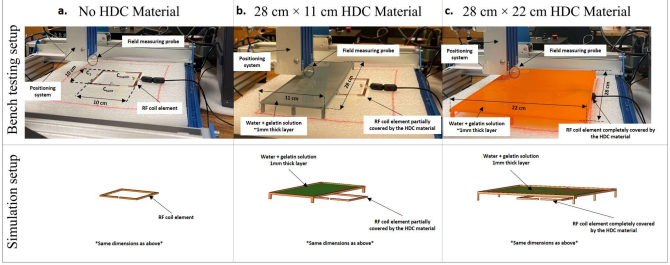

Figure 5. Bench test measurement setup and the identical simulation setup used for verification. (a) The high-impedance coil without any high dielectric constant material placed on top kept on the low loss platform on the 3D positioning system and field measuring probe. (b)The high-impedance coil and the 3D printed tray containing the water + gelatin solution mimicking one of the high relative permittivity values partially covering the high impedance coil below. The partially covering 3D printed tray had the dimensions of 11×28 cm$^2$ and the distance between the high dielectric constant material from the high-impedance coil was approximately 10cm. (c)The high impedance coil and 22×28 cm$^2$ 3D printed tray with water + gelatin solution added on top. The distance between the high dielectric constant material and the high-impedance coil was approximately 10cm.

III. RESULTS

A. Inter-element Isolation

The scattering parameters were used to calculate the isolation values between the high-impedance coils. **Figures 2**, **3**, and **4** depict the simulation setup for each scenario under consideration. Following the previous procedure, the high-impedance coils were separated by 1 cm. The only modifications in the design were related to the relative permittivity of the HDC material used and the distance from the coil while maintaining a constant distance between the phantom and high-impedance coils in each scenario. With no HDC material present, the inter-element isolation value of the high-impedance coils was -25.57 dB. The S-parameter linear plot for high-impedance coils without HDC material is shown in **Fig. 6**. We used this value as the gold standard to compare the inter-element isolation values of different circumstances involving the application of HDC material to determine the effect the HDC material had on the isolation performance. The inter-element isolation values for high-impedance coils with HDC material above them were also evaluated. **Fig. 7.** depicts the scattering parameters for the high-impedance coils, with a 30 × 11.5 cm$^2$ HDC material sheet partially covering them. The simulation setup shown in **Fig. 3** was used for the evaluations. The HDC material was placed between the high-impedance coils at distances ranging from 3 mm to 10 mm, with 3 mm being the closest. The HDC material had relative permittivity values of 50 and 100 at a distance of 3 mm, with inter-element isolation values of -23.9 dB and -23.1 dB, respectively. The HDC material with relative permittivity values of 50 and 100 was used for a 5 mm distance to achieve inter-element isolation values of -23.9 dB and -23.2 dB, respectively. Furthermore, the relative permittivity values for the HDC material were 50, 100, and 150 for an 8 mm distance, and the inter-element isolation values were -23.9 dB, -23.5 dB, and -23 dB, respectively. For the longest distance tested, 10 mm, the HDC material had relative permittivity values of 50, 100, 150, and 200, and the inter-element isolation values were -24 dB, -22.9 dB, -23.4 dB, and -23.3 dB, respectively. The inter-element isolation for each variable case tested for the 30 × 11.5 cm$^2$ HDC material sheet partially covering the high-impedance coils was at least -20 dB, preserving the high-impedance coils' inter-element isolation performance. The scattering parameters for the other 30 × 23 cm² HDC material sheet that completely covered the high-impedance coils are shown in **Fig. 8**. As with the previous HDC material sheet, the distance between the high-impedance coils on the given HDC material sheet was increased from 3 mm to 10 mm. For each distance tested, different relative permittivity values were used to validate the inter-element isolation performance of the high-impedance coils. Relative permittivity values of 50 and 100 were used for a distance of 3mm between the HDC material and the high-impedance coils, with inter-element isolation values of -21.8 dB and -20.7 dB, respectively. The HDC material with relative permittivity of 50 and 100 was also used for the 5mm distance, with corresponding inter-element isolation values of -21.7 dB and -20.8 dB. At a distance of 8 mm from the high-impedance coils, the HDC material was evaluated using relative permittivity values of 50, 100, and 150. In that order, the inter-element isolation values were -22.3 dB, -21 dB, and -20 dB. The relative permittivity values used were 50, 100, 150, and 200, with corresponding inter-element isolation values of -22.8 dB, -21.7 dB, -20.8 dB, and -20.4 dB for the 10 mm distance. The use of the HIC material produced inter-element isolation values of -20 dB or better between the high-impedance coils used in both evaluated cases, preserving the coils' isolation performance.

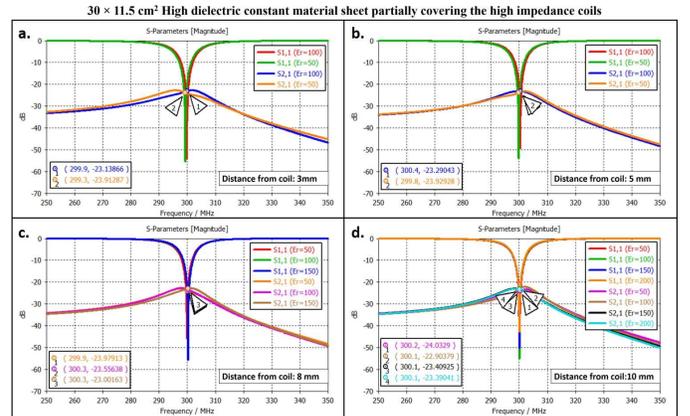

Figure 7. Scattering parameters for the experimental cases involving the 30 × 11.5 cm$^2$ HDC material sheet partially covering the high-impedance coils. The figures compile the cases based on the distance of the HDC material from the high-impedance coils. (a) The S-parameters for the cases involving the HDC material placed 3mm away from the high-impedance coils. (b) The S-parameters for the cases involving the HDC material placed 5mm away from the high-impedance coils. (c) The S-parameters for the cases involving the HDC material placed 8mm away from the high impedance coils. (d) The S-parameters for the cases involving the HDC material placed 10mm away from the high-impedance coils.

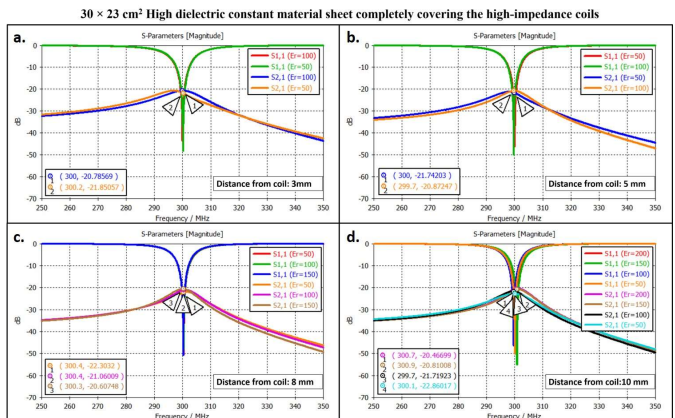

Figure 8. Scattering parameters for the experimental cases involving the 30 × 23 cm² HDC material sheet completely covering the high-impedance coils. The figures compile the cases based on the distance of the HDC material from the high-impedance coils. (a) The S-parameters for the cases involving the HDC material placed 3mm away from the high-impedance coils. (b) The S-parameters for the cases involving the HDC material placed 5mm away from the high-impedance coils. (c) The S-parameters for the cases involving the HDC material placed 8mm away from the high impedance coils. (d) The S-parameters for the cases involving the HDC material placed 10mm away from the high-impedance coils.

### B. Electric fields

To assess the effect of HDC material on electric field values and distribution over the phantom, electric fields were computed using 3D electromagnetic simulations. A cylindrical phantom was included in the simulation setup to facilitate this, and different human tissue material properties were assigned to the phantom to investigate the electric field behavior of the high-impedance coils with the HDC material for various human tissue properties. The cylindrical phantom's tissue properties included values for the human brain, kidney, breast fat, and tendon/ligament. The mentioned human tissue samples were chosen to facilitate the evaluation of various relative permittivity and electrical conductivity values. The material properties of the Brain phantom were as follows: relative permittivity $\varepsilon_r$: 50, electrical conductivity 0.6 S/m, and density

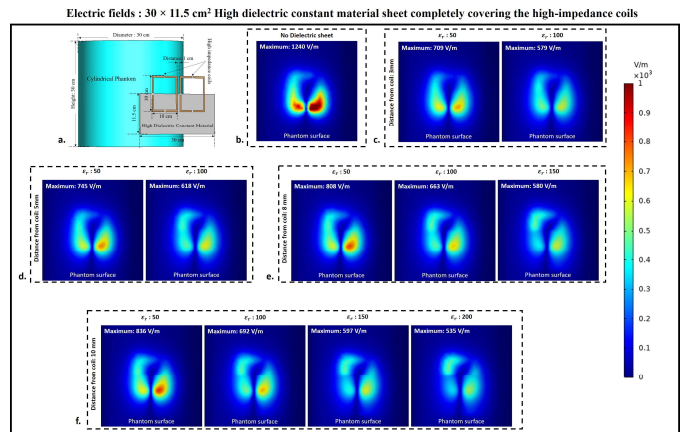

Figure 9. Electric field distribution on the surface of the cylindrical phantom used in the experimental cases involving 30×11.5 cm² HDC material placed in between the phantom and the high-impedance coil. (a) The simulation setup used for the electric field evaluation. (b) HDC material-free high-impedance coil E-field distribution. (c) E-field distribution for HDC material 3mm from high-impedance coils. (d) E-field distribution for HDC material 5mm from high-impedance coils. (e) E-field distribution for HDC material 8mm from high-impedance coils. (f) E-field distribution for HDC material 10mm from high-impedance coils.

1000 Kg/m³. The Kidney phantom's material properties were as follows: relative permittivity $\varepsilon_r$: 70.5, electrical conductivity 1.02 S/m, and density 1066 Kg/m³. The material properties of the Breast fat phantom were as follows: relative permittivity $\varepsilon_r$: 5.54, electrical conductivity: 0.0327 S/m, and density: 911 Kg/m³.Finally, the Tendon/Ligament phantom material properties were as follows: relative permittivity $\varepsilon_r$: 48, electrical conductivity 0.537 S/m, and density 1142 Kg/m³. The electric field distribution on the surface of the cylindrical phantom assigned with human brain tissue properties for the high-impedance coils covered with HDC materials of 30 × 11.5 cm² dimension is shown in **Fig.9**. When placed without any HDC material between them and the phantom, the high-impedance coils produced a peak electric field value of 1240 V/m on the phantom surface. The peak electric field values for the HDC materials kept 3 mm away from the coils were 709 V/m and 579 V/m for the relative permittivity of 50 and 100, respectively. Furthermore, for a 5mm distance between the coil and the HDC material, the peak electric field values of 745 V/m and 618 V/m for the relative permittivity of 50 and 100, respectively, were obtained. When the HDC material was 8 mm away from the high-impedance coils, the peak electric field values were 808 V/m, 663 V/m, and 580 V/m for the relative permittivity of 50, 100, and 150, respectively. Finally, for the largest distance evaluated (10 mm), peak electric fields of 836 V/m, 692 V/m, 597 V/m, and 535 V/m were observed for relative permittivity of 50,100,150, and 200, respectively. As a result, when HDC material with a relative permittivity of 200 was used 10 mm away from the high-impedance coils, the lowest peak electric field values (535 V/m) were observed. **Fig. 10** depicts the electric field distribution on the surface of the cylindrical phantom for high-impedance coils made of HDC materials with dimensions of 30 × 23 cm². The human brain material properties were assigned to the cylindrical phantom for electric field distribution evaluation, as in the previous case. The peak electric field value for high-impedance coils without HDC remains constant at 1240 V/m. Peak electric field values of 659 V/m for the relative permittivity of 50 and 511 V/m for the relative permittivity of 100 were observed when the HDC material was 3 mm away from the high-impedance coils. Peak electric field values of 730 V/m for the relative permittivity of 50 and 570 V/m for the relative permittivity of 100 were

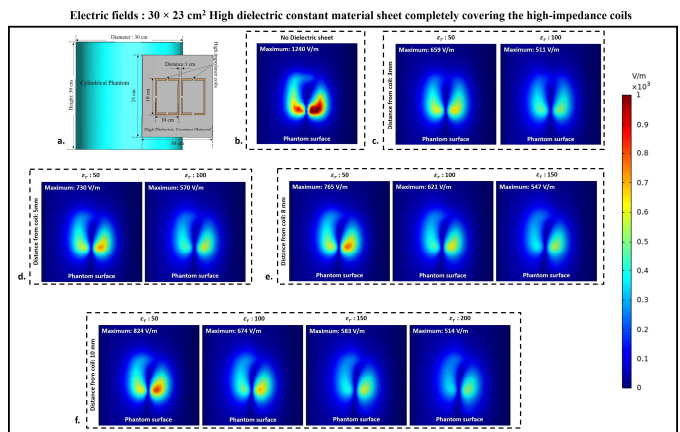

Figure 10. Electric field distribution on the surface of the cylindrical phantom used in the experimental cases involving 30×23 cm² HDC material placed in between the phantom and the high-impedance coil. (a) The simulation setup used for the electric field evaluation. (b) HDC material-free high-impedance coil E-field distribution. (c) E-field distribution for HDC material 3mm from high-impedance coils. (d) E-field distribution for HDC material 5mm from high-impedance coils. (e) E-field distribution for HDC material 8mm from high-impedance coils. (f) E-field distribution for HDC material 10mm from high-impedance coils.

observed for the HDC material placed 5 mm away from the high-impedance coils. Furthermore, the peak electric field value was 765 V/m for the relative permittivity of 50, 621 V/m for the relative permittivity of 100, and 547 V/m for the relative permittivity of 150 for an 8 mm distance between the material and the coil. Finally, the peak electric field value was 824 V/m for the relative permittivity of 50, 674 V/m for the relative permittivity of 100, 583 V/m for the relative permittivity of 150, and 514 V/m for the relative permittivity of 200 for the farthest distance evaluated, which was 10 mm. As a result, when the HDC material with a relative permittivity of 100 was kept 3 mm away from the coil, the lowest peak electric field value of 511 V/m was observed. Other human tissue properties, such as kidney, breast fat, and tendon/ligament, were assigned to the cylindrical phantom, and similar cases were evaluated to obtain peak electric field values on the phantom surface. **Fig. 11** depicts one-dimensional profiles for the peak electric field trend for each human tissue parameter evaluated. For each tissue property, the peak electric field values on the phantom surface decrease in the same pattern. When a partially covering HDC material with dimensions of 30 × 11.5 cm² was used and kept at a distance of 10 mm from the high impedance coils, the maximum reduction in peak electric field values was observed for all of the evaluated tissue properties. Furthermore, when another topology of HDC material with dimensions of 30 × 23 cm² was used, the brain, kidney, and tendon/ligament phantoms

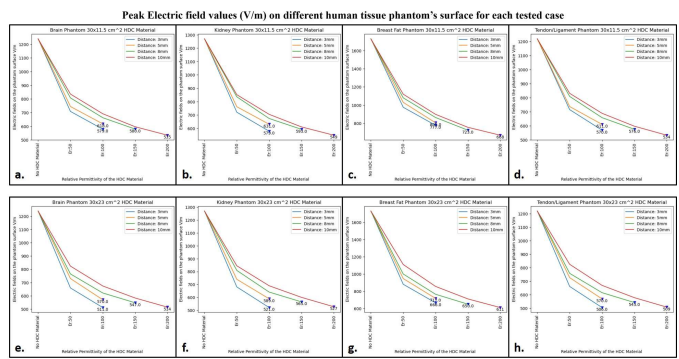

Figure 11. The 1D profiles of the peak electric field strengths evaluated for various cases involving different topology of the HDC material, its distance from the high-impedance coils, its relative permittivity and various human tissue properties assigned to the cylindrical phantom. (a) 30× 11.5 cm² HDC material: Brain Phantom (b) 30× 11.5 cm² HDC material: Kidney Phantom (c) 30× 11.5 cm² HDC material: Breast Fat Phantom (d) 30× 11.5 cm² HDC material: Tendon/ligament Phantom (e) 30× 23 cm² HDC material: Brain Phantom (f) 30× 23 cm² HDC material: Kidney Phantom (g) 30× 23 cm² HDC material: Breast Fat Phantom (h) 30× 23 cm² HDC material: Tendon/ligament Phantom

had a significant decrease in peak electric field strength when HDC material with a relative permittivity of 100 was used, and kept 3 mm away from the coils. In addition, for the Breast fat phantom, the HDC material with a relative permittivity of 200 and a distance of 10 mm from the coils showed a significant reduction in peak electric fields on the phantom surface. The peak electric field values were reduced by 56.85% when a partially covering high dielectric constant material measuring 30×11.5 cm² was used. Additionally, when high dielectric constant materials measuring 30×23 cm² were used to completely cover the RF coils, the peak electric field values were reduced by 58.54%. These reductions were observed in a phantom with properties assigned to mimic human brain tissue. The 30×11.5 cm² high dielectric constant material sheet resulted in a reduction of the peak electric field values for the kidney, breast fat, and tendon/ligament by 57.16%, 61.38%, and 56.22%, respectively. Furthermore, kidney, breast fat, and tendon/ligament tissue properties showed a 58.97%, 64.68%, and 58.27% reduction in peak electric field values when the RF coils were fully covered by a 30×23 cm² high dielectric constant material sheet.

### C. Specific Absorption rate

Using electromagnetic simulations, the peak specific absorption rate values (W/Kg) in the cylindrical phantom assigned with human tissue properties were calculated. The cylindrical phantom used in the simulations was assigned similar human tissue properties such as the human brain, kidney, breast fat, and tendon/ligament, and the SAR distribution and peak SAR values were recorded for all of the stated tissue properties. The SAR distribution on the cylindrical phantom assigned the human brain tissue properties, as well as the peak SAR values for the respective field distribution for the high impedance coils covered with 30 × 11.5 cm² HDC material, are shown in **Fig.12**. The peak SAR value of the high-impedance coils without HDC materials was 4.45 W/kg. The evaluated cases for the HDC material with dimensions of 30 × 11.5 cm² and a distance from the high impedance coils ranging from 3mm to 10mm, on the other hand, had the following peak

SAR values: Peak SAR values of 4.44 and 4.41 W/Kg were observed for the material with relative permittivity of 50 and

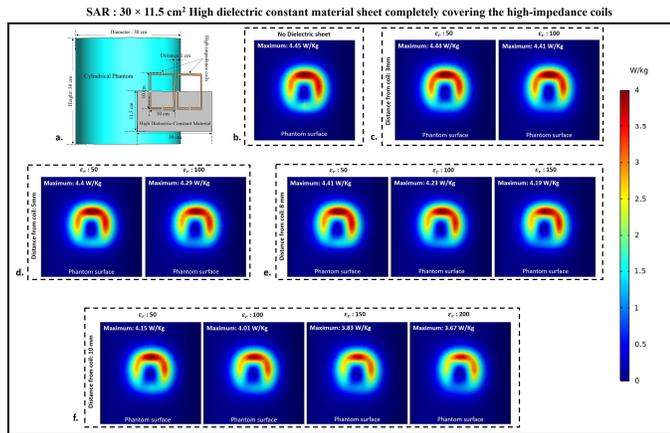

Figure 12. Specific Absorption Rate (SAR) distribution on the surface of the cylindrical phantom used in the experimental cases involving 30×11.5 cm² HDC material placed in between the phantom and the high-impedance coil. (a) The simulation setup used for the SAR evaluation. (b) HDC material-free high-impedance coil SAR distribution. (c) SAR distribution for HDC material 3mm from high-impedance coils. (d) SAR distribution for HDC material 5mm from high-impedance coils. (e) SAR distribution for HDC material 8mm from high-impedance coils. (f) SAR distribution for HDC material 10mm from high-impedance coils.

100, respectively, for an evaluated distance of 3mm between the HDC material and the high impedance coils. Similarly, for the material with relative permittivity of 50 and 100, peak SAR values of 4.4 and 4.29 W/Kg were observed for the 5 mm evaluated distance between the material and the coils, respectively. Furthermore, for a distance of 8mm, the peak SAR values for the material with relative permittivity of 50,100, and 150 were 4.41,4.23, and 4.19 W/Kg, respectively. Finally, for the material with relative permittivity of 50,100,150, and 200, peak SAR values of 4.15,4.01,3.83,3.67 W/Kg were observed for the evaluated distance of 10mm. Similar cases were evaluated for the HDC material with different dimensions of 30 × 23 cm² following the same pattern. **Fig.13.** depicts the SAR field distribution, as well as the peak SAR values recorded for each evaluated case using HDC material with dimensions of 30 × 23 cm². Peak SAR values of 4.49 and 4.48 W/Kg were observed for the material with relative permittivity of 50 and 100 at a 3mm distance between the material and the high-impedance coils, respectively. Similarly, for a 5mm distance, the peak SAR values were 4.47 and 4.39 W/Kg for materials with relative permittivity of 50 and 100, respectively. Furthermore, for the 8mm distance, peak SAR values of 4.4,4.3,4.25 W/Kg were obtained for materials with relative permittivity values of 50, 100, and 150, respectively. Finally, for the material with relative permittivity of 50,100,150, and 200, peak SAR values of 4.24,4.12,4, and 3.84 W/Kg were recorded for the maximum distance evaluated of 10mm. Peak SAR values for other human tissue properties such as kidney, breast fat, and tendon/ligament were also evaluated for cylindrical phantoms. **Fig. 14** depicts the 1D profiles of peak SAR values for all of the cases studied, which involved HDC material sheets with varying relative permittivity values and varying distances between them and the high-impedance coils. The selected human tissue properties include a wide range of relative permittivity and electric conductivity values found inside the human body and will provide additional insights into SAR value behavior based on relative permittivity and electric

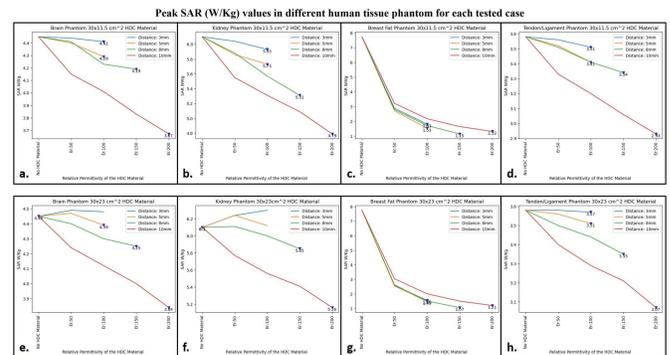

Figure 14. The 1D profiles of the peak SAR values evaluated for various cases involving different topology of the HDC material, its distance from the high-impedance coils, its relative permittivity and various human tissue properties assigned to the cylindrical phantom. (a) 30× 11.5 cm² HDC material: Brain Phantom (b) 30× 11.5 cm² HDC material: Kidney Phantom (c) 30× 11.5 cm² HDC material: Breast Fat Phantom (d) 30× 11.5 cm² HDC material: Tendon/ligament Phantom (e) 30× 23 cm² HDC material: Brain Phantom (f) 30× 23 cm² HDC material: Kidney Phantom (g) 30× 23 cm² HDC material: Breast Fat Phantom (h) 30× 23 cm² HDC material: Tendon/ligament Phantom

conductivity values. When placed without any HDC material in between, the high impedance coils deposited peak SAR values of 4.45 W/Kg, 6.1 W/Kg, 7.74 W/Kg, and 3.58 W/Kg in the brain, kidney, breast fat, and tendon/ligament phantoms, respectively. After inserting the HDC material with dimensions of 30 × 11.5 cm² between the phantom and the coils, the peak SAR value in the brain phantom was reduced to 3.67 W/Kg, 4.79 W/Kg in the kidney, 1.32 W/Kg in the Breast fat phantom, and 2.93 W/Kg in the Tendon/ligament phantom. In a similar pattern, for the HDC material with dimensions of 30 × 23 cm², the peak SAR value for the brain phantom was reduced to 3.84 W/Kg, the peak SAR value for the kidney phantom was reduced to 5.16 W/Kg, the peak SAR value for the breast fat was reduced to 1.21 W/Kg, and the peak SAR value for the tendon/ligament was reduced to 3.07 W/Kg.

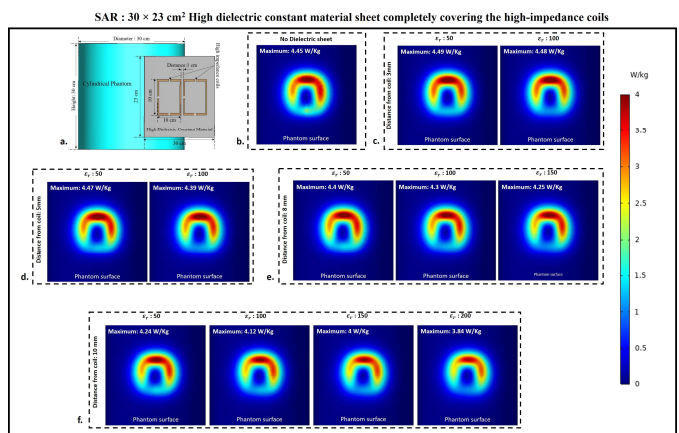

Figure 13. Specific Absorption Rate (SAR) distribution on the surface of the cylindrical phantom used in the experimental cases involving 30×23 cm² HDC material placed in between the phantom and the high-impedance coil. (a) The simulation setup used for the SAR evaluation. (b) HDC material-free high-impedance coil SAR distribution. (c) SAR distribution for HDC material 3mm from high-impedance coils. (d) SAR distribution for HDC material 5mm from high-impedance coils. (e) SAR distribution for HDC material 8mm from high-impedance coils. (f) SAR distribution for HDC material 10mm from high-impedance coils.

The utilization of high dielectric constant material, with dimensions of 30×11.5 cm$^2$, resulted in a reduction of peak specific absorption rate (SAR) values by 17.52%. Similarly, when the high dielectric constant materials, with dimensions of 30×23 cm$^2$, completely covered the RF coils, a reduction of peak SAR values by 13.70% was observed. These evaluations were conducted on a phantom model that simulated human brain tissue properties. The high dielectric constant material sheet with dimensions of 30×11.5 cm$^2$ resulted in peak SAR value reductions of 21.47%, 85.14%, and 18.15% in kidney, breast fat, and tendon/ligament tissue properties, respectively. Furthermore, the peak SAR values for kidney, breast fat, and tendon/ligament tissue properties were reduced by 15.40%, 86.17%, and 14.24%, respectively, by the 30×23 cm$^2$ high dielectric constant material sheet that covered the entire RF coils.

In each analyzed scenario. Peak B1 values of 3.9 $\mu T/\sqrt{W}$ were generated by the high-impedance coils devoid of HDC material. It was intended to compare this beginning value to the peak B1 field values computed for each example using HDC material with a particular relative permittivity value to ascertain whether the HDC material had any adverse effects on the B1 field distribution of the high-impedance coils. Similar cases involving the use of HDC material with different relative permittivity values and distance from the high-impedance coils were evaluated for the B1 field distribution and peak B1 field strength in the cylindrical phantom's central sagittal slice. For the HDC material kept 3mm away, two relative permittivity values of 50 and 100 were used, yielding peak B1 field strengths of 3.97 and 3.97 $\mu T/\sqrt{W}$, respectively. Similarly, for the HDC material 5mm away from the high-impedance coils, the same relative permittivity values (50 and 100) produced peak B1 field strengths of 4.02 and 4.04 $\mu T/\sqrt{W}$, respectively. Furthermore, for the HDC material placed 8mm away, relative permittivity

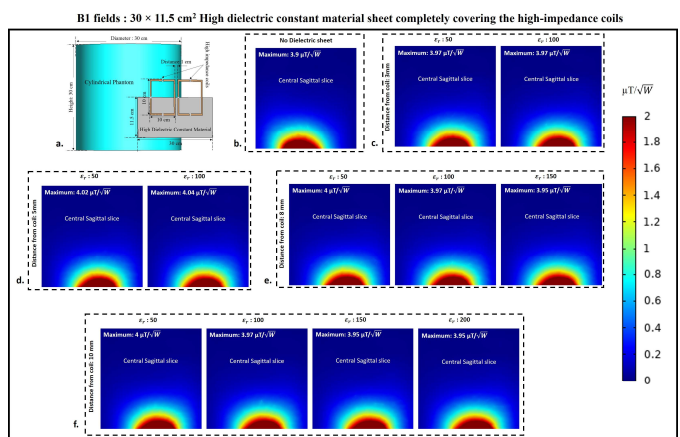

Figure 15. B1 field distribution on the surface of the cylindrical phantom used in the experimental cases involving 30×11.5 cm$^2$ HDC material placed in between the phantom and the high-impedance coil. (a) The simulation setup used for the B1 field evaluation. (b) HDC material-free high-impedance coil B1-field distribution. (c) E-field distribution for HDC material 3mm from high-impedance coils. (d) B1-field distribution for HDC material 5mm from high-impedance coils. (e) B1-field distribution for HDC material 8mm from high-impedance coils. (f) B1-field distribution for HDC material 10mm from high-impedance coils.

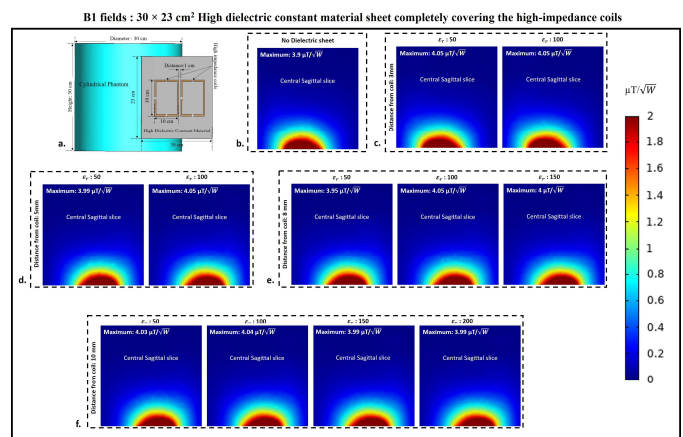

Figure 16. B1 field distribution on the surface of the cylindrical phantom used in the experimental cases involving 30×23 cm$^2$ HDC material placed in between the phantom and the high-impedance coil. (a) The simulation setup used for the B1 field evaluation. (b) HDC material-free high-impedance coil B1-field distribution. (c) E-field distribution for HDC material 3mm from high-impedance coils. (d) B1-field distribution for HDC material 5mm from high-impedance coils. (e) B1-field distribution for HDC material 8mm from high-impedance coils. (f) B1-field distribution for HDC material 10mm from high-impedance coils.

### D. B1 field distribution

B1 field distribution in the central sagittal slice of the cylindrical phantom assigned with human brain tissue properties was computed to assess how the presence of HDC material sheet between the phantom and the high-impedance coils affects the B1 field efficiency of the high-impedance coils. **Fig. 15** displays the high-impedance coils' B1 field distribution in the center sagittal slice of the cylindrical phantom, which is partially covered by a sheet of HDC material with measurements of 30 × 11.5 cm$^2$. Each analyzed case's peak B1 field values were also assessed in the central sagittal slice to track the pattern of the peak B1 field strength with its distribution. Without using any HDC material, the high-impedance coils' initial B1 field distribution and peak B1 value were calculated. For a fair comparison, the computed B1 field strength values in µT were normalized by dividing by the square root of the accepted power for the high-impedance coils values of 50,100, and 150 were used, yielding peak B1 field strengths of 4, 3.97, and 3.95 $\mu T/\sqrt{W}$, respectively. Finally, at a distance of 10mm between the HDC material and the high-impedance coils, relative permittivity values of 50,100,150, and 200 were used for the HDC material, with peak B1 field strengths of 4,3.97,3.95, and 3.95 $\mu T/\sqrt{W}$, respectively. The evaluated cases involving HDC material of dimensions 30×11.5 cm$^2$ consistently produced peak B1 field strengths greater than the initial recorded value of 3.9 $\mu T/\sqrt{W}$ and preserved the high-impedance coil structure's B1 field distribution. Similarly, for the experimental cases involving the use of HDC material with dimensions 30×23 cm$^2$, the B1 field distribution in the sagittal slice of the cylindrical phantom assigned with human brain tissue properties was evaluated. The B1 field distribution for all of the evaluated cases for the HDC material with the specified dimensions is shown in **Fig. 16**. For a distance of 3mm between the HDC material and the high-impedance coils, the HDC material was assigned two relative permittivity values of 50 and

100, and peak B1 field strengths of 4.05 and 4.05 $\mu T/\sqrt{W}$ were observed in the phantom's central sagittal slice, respectively. Relative permittivity values of 50 and 100 were used for the evaluated distance of 5mm, and peak B1 field strengths of 3.99 and 4.05 $\mu T/\sqrt{W}$ were recorded, respectively. Furthermore, for the 8mm distance, the HDC material had relative permittivity values of 50,100, and 150, with peak B1 field strengths of 3.95, 4.05, and 4 $\mu T/\sqrt{W}$, respectively. Finally, for a 10mm distance, the HDC material had relative permittivity values of 50,100,150, and 200, with peak B1 field strengths of 4.03, 4.04, 3.99, and 3.99 $\mu T/\sqrt{W}$, respectively.

### E. Bench Test Results

The simulated results were validated using bench test results using the experimental setup shown in **Fig.5.** For all of the evaluated cases, the field distributions were plotted on an axial plane approximately 15 mm from the coil, and no phantom was used in simulations or bench tests. The measured and simulated electric field distributions for the three evaluated cases are shown in **Fig. 17.** High-impedance coils with no HDC material present, high-impedance coils with a 28×11 cm² 3D printed tray with a 1mm thick HDC material layer added on top, and high-impedance coils with a 28×22 cm² 3D printed tray with a 1mm thick HDC material layer added on top are among the cases. For each case, the field distributions were plotted in an axial plane 15mm away from the coil. In addition to the field distribution, the peak electric field strength was measured for each of the cases studied. In the bench testing setup, the first evaluated case involving high-impedance coils without any HDC material produced a peak electric field strength of 1254.11 V/m versus 1691 V/m in the simulation setup. Furthermore, in the bench testing setup, a high-impedance coil covered by the HDC material tray with dimensions of 28×11 cm² produced a peak electric field strength of 586.66 V/m vs. 710 V/m in the simulation setup. Finally, in the bench testing setup, a high-impedance coil covered by the HDC material tray with dimensions of 28×22 cm² produced a peak electric field strength of 596.1 V/m vs. 656 V/m in the simulation setup. Overall, the measured peak electric field value for the bench testing setup was reduced by 53.22% when HDC material added to the 3D printed tray with dimensions of 28×11 cm² was placed above the high impedance coil, and it was reduced by 52.46% when HDC material added to the 3D printed tray with different dimensions of 28×22 cm² was placed above the high impedance coil. In identical simulation setup, The peak electric field strength was reduced by 58.01% with the application of HDC material with dimensions of 28×11 cm² and by 61.20% with the addition of HDC material with dimensions of 28×22 cm² above the high-impedance coil.

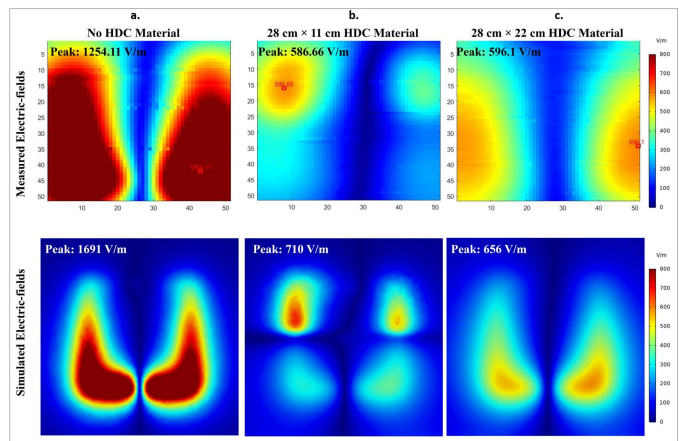

Figure 17. The measured electric field distribution on bench Vs. simulated electric field distribution in an identical setup. (a) HDC material free high impedane coil measured vs simulated E-field distribution (b) The high-impedance coil with 28×11cm² HDC material sheet measured vs simulated E-field distribution (c) The high-impedance coil with 28×22cm² HDC material sheet measured vs simulated E-field distribution.

Magnetic field distribution in $\mu T/\sqrt{W}$ was compared between the experimental and simulation setups to further validate the concept. The B1 field distribution was evaluated in both the experimental bench testing setup and the simulation setup for three cases, including high-impedance coils with and without the HDC material placed on top of the coils (see **Fig. 18**). The measured and simulated results agreed well, with an improved overall B1 field strength for the case involving the application of HDC material with dimensions of 28×11 cm² and a slightly reduced B1 field strength for the case involving the application of HDC material with dimensions of 28×22 cm². The disparity between measured and simulated values can be attributed to a difference in the accuracy of the results between bench testing and simulation setup. The measured field distributions were reconstructed by measuring each field on an axial plane about 15 mm away from the coil. The measurement grid included a 51×51 matrix with a precision of 2mm, and the simulation setup included very fine mesh settings with a specified minimum element length of 0.1mm. The higher simulation accuracy resulted in overall higher field strength values. Another factor that contributes to the discrepancy is the relative permittivity of the HDC material used. The HDC material was created by combining gelatin and distilled water, and its relative permittivity was estimated to be around 78. However, the practical value of the HDC material could be higher or lower than the assumed value, resulting in higher peak electric field values in the simulated results.

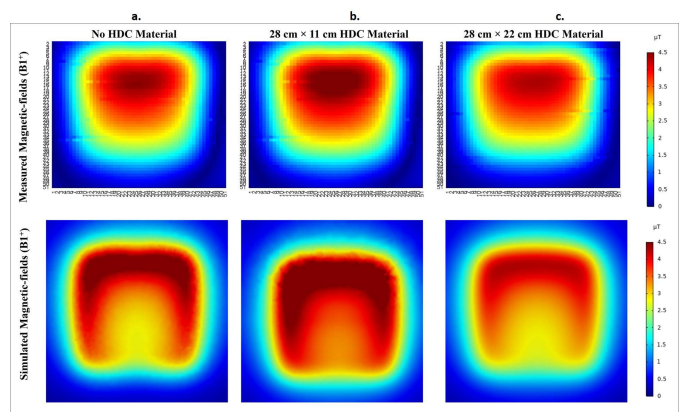

Figure 18. The measured B1 field distribution on bench Vs. simulated B1 field distribution in an identical setup. (a) HDC material free high impedane coil measured vs simulated B1-field distribution (b) The high-impedance coil with 28×11cm² HDC material sheet measured vs simulated B1-field distribution (c) The high-impedance coil with 28×22cm² HDC material sheet measured vs simulated B1-field distribution.

## IV. Conclusions and Discussions

In this work, we numerically and experimentally evaluated a novel method of incorporating high dielectric constant material to reduce electric fields and SAR values in high-impedance RF coils and arrays. Using numerical simulations, experimental cases introducing high dielectric constant material to high-impedance coils and strategic placement of the high dielectric constant material with appropriate relative permittivity to maximize E-field and SAR reduction were evaluated to demonstrate the proposed method's success in lowering electric fields and SAR values over high-impedance coils. When compared to high dielectric constant material free high impedance coils, our proposed approach successfully reduced the peak electric field values by at least 50% and the SAR values by at least 13% in numerical simulations while preserving the initial B1 efficiency and decoupling performance of the high impedance coils. Following the successful evaluation of the proposed method in numerical simulations, a prototype was tested on the bench to test one of the experimental setups, and the results demonstrated preservation of B1 efficiency and peak electric field value reduction of at least 50%, validating the simulation results. Our findings demonstrate the potential of high dielectric constant materials in ultra-high field MR imaging as a viable option for alleviating electromagnetic exposure concerns and enabling safer MR imaging at ultra-high fields. The use of high-dielectric constant materials for exposure reduction is not limited to high-impedance coils, but can also be used in conjunction with other RF coils. Further research into the use of high dielectric constant materials in MRI RF hardware will undoubtedly contribute to safer and more efficient electromagnetic systems.

## Acknowledgement

This work is supported in part by the NIH under a BRP grant U01 EB023829 and by the State University of New York (SUNY) under SUNY Empire Innovation Professorship Award.

## References

[1] Hayes CE. The development of the birdcage resonator: a historical perspective. NMR Biomed 2009;22(9):908-18.
[2] Zhang X, Ugurbil K, Sainati R, Chen W. An inverted-microstrip resonator for human head proton MR imaging at 7 tesla. IEEE Trans Biomed Eng 2005;52(3):495-504.
[3] Zhang X, Zhu X, Qiao H, Liu H, Vaughan JT, Ugurbil K, et al. A circular-polarized double-tuned (31P and 1H) TEM coil for human head MRI/MRS at 7T. *Proc. Intl. Soc. Mag. Reson. Med.* 2003:423.
[4] Pang Y, Wong EW, Yu B, Zhang X. Design and numerical evaluation of a volume coil array for parallel MR imaging at ultrahigh fields. Quant Imaging Med Surg 2014;4(1):50-6.
[5] Wu B, Wang C, Kelley DA, Xu D, Vigneron DB, Nelson SJ, et al. Shielded microstrip array for 7T human MR imaging. IEEE Trans Med Imaging 2010;29(1):179-84.
[6] Wu B, Zhang X, Wang C, Li Y, Pang Y, Lu J, et al. Flexible transceiver array for ultrahigh field human MR imaging. Magn Reson Med 2012;68(4):1332-8.
[7] Pang Y, Xie Z, Li Y, Xu D, Vigneron D, Zhang X. Resonant Mode Reduction in Radiofrequency Volume Coils for Ultrahigh Field Magnetic Resonance Imaging. Materials (Basel) 2011;4(8):1333-44.
[8] Zhang X, Ugurbil K, Chen W. Microstrip RF surface coil design for extremely high-field MRI and spectroscopy. Magn Reson Med 2001;46(3):443-50.
[9] Zhang X, Ugurbil K, Chen W. A microstrip transmission line volume coil for human head MR imaging at 4T. J Magn Reson 2003;161(2):242-51.
[10] Wu B, Wang C, Lu J, Pang Y, Nelson SJ, Vigneron DB, et al. Multi-channel microstrip transceiver arrays using harmonics for high field MR imaging in humans. IEEE Trans Med Imaging 2012;31(2):183-91.
[11] Zhang X, Burr AR, Zhu X, Adriany G, Ugurbil K, Chen W. A Dual-tuned Microstrip Volume Coil Array for Human Head parallel 1H/31P MRI/MRS at 7T. *the 11th Scientific Meeting and Exhibition of ISMRM.* Miami, Florida; 2005:896.
[12] Pang Y, Wu B, Wang C, Vigneron DB, Zhang X. Numerical Analysis of Human Sample Effect on RF Penetration and Liver MR Imaging at Ultrahigh Field. Concepts Magn Reson Part B Magn Reson Eng 2011;39B(4):206-16.
[13] Adriany G, Van de Moortele PF, Wiesinger F, Moeller S, Strupp JP, Andersen P, et al. Transmit and receive transmission line arrays for 7 Tesla parallel imaging. Magn Reson Med 2005;53(2):434-45.
[14] Zhang X, Ugurbil K, Chen W. Method and apparatus for magnetic resonance imaging and spectroscopy using microstrip transmission line coils. 7023209. US patent 2006.
[15] Yan X, Xue R, Zhang X. Closely-spaced double-row microstrip RF arrays for parallel MR imaging at ultrahigh fields. Appl Magn Reson 2015;46(11):1239-48.
[16] Zhang X, Pang Y. Parallel Excitation in Ultrahigh Field Human MR Imaging and Multi-Channel Transmit System. OMICS J Radiol 2012;1(3):e110.
[17] Yan X, Zhang X. Decoupling and matching network for monopole antenna arrays in ultrahigh field MRI. Quant Imaging Med Surg 2015;5(4):546-51.
[18] Zhang X. Method and apparatus for MRI signal excitation and reception using non-resonance RF method (NORM). USA: The Regents of The University of California, Oakland, CA; 2012.
[19] Pang Y, Zhang X, Xie Z, Wang C, Vigneron DB. Common-mode differential-mode (CMDM) method for double-nuclear MR signal excitation and reception at ultrahigh fields. IEEE Trans Med Imaging 2011;30(11):1965-73.

[20] Pang Y, Zhang X. Precompensation for mutual coupling between array elements in parallel excitation. Quant Imaging Med Surg 2011;1(1):4-10.

[21] Rutledge O, Kwak T, Cao P, Zhang X. Design and test of a double-nuclear RF coil for (1)H MRI and (13)C MRSI at 7T. J Magn Reson 2016;267:15-21.

[22] Payne K, Zhao Y, Ying LL, Zhang X. Design of a well decoupled 4-channel catheter Radio Frequency coil array for endovascular MR imaging at 3T. Proc Int Soc Magn Reson Med Sci Meet Exhib Int Soc Magn Reson Med Sci Meet Exhib 2023;31.

[23] Yan X, Wei L, Xue R, Zhang X. Hybrid monopole/loop coil array for human head MR imaging at 7T. Appl Magn Reson 2015;46(5):541-50.

[24] Yan X, Xue R, Zhang X. A monopole/loop dual-tuned RF coil for ultrahigh field MRI. Quant Imaging Med Surg 2014;4(4):225-31.

[25] Avdievich NI, Pan JW, Hetherington HP. Resonant inductive decoupling (RID) for transceiver arrays to compensate for both reactive and resistive components of the mutual impedance. NMR Biomed 2013;26(11):1547-54.

[26] Roemer PB, Edelstein WA, Hayes CE, Souza SP, Mueller OM. The NMR phased array. Magn Reson Med 1990;16(2):192-225.

[27] Wu B, Wang C, Krug R, Kelley DA, Xu D, Pang Y, et al. 7T human spine imaging arrays with adjustable inductive decoupling. IEEE Trans Biomed Eng 2010;57(2):397-403.

[28] Wu B, Zhang X, Qu P, Shen GX. Design of an inductively decoupled microstrip array at 9.4 T. J Magn Reson 2006;182(1):126-32.

[29] Wu B, Zhang X, Qu P, Shen GX. Capacitively decoupled tunable loop microstrip (TLM) array at 7 T. Magn Reson Imaging 2007;25(3):418-24.

[30] Xie Z, Zhang X. An 8-channel microstrip array coil for mouse parallel MR imaging at 7T by using magnetic wall decoupling technique. In: ISMRM PottAMo, ed. *the 16th Annual Meeting of ISMRM.* Toronto, Canada; 2008:2973.

[31] Xie Z, Zhang X. A novel decoupling technique for non-overlapped microstrip array coil at 7T MR imaging. In: ISMRM PottAMo, ed. *the 16th Annual Meeting of ISMRM.* Toronto, Canada; 2008:1068.

[32] Xie Z, Zhang X. An 8-channel non-overlapped spinal cord array coil for 7T MR imaging. In: ISMRM PottAMo, ed. *the 16th Annual Meeting of ISMRM.* Toronto, Canada; 2008:2974.

[33] Xie Z, Zhang X. An eigenvalue/eigenvector analysis of decoupling methods and its application at 7T MR imaging. *the 16th Annual Meeting of ISMRM Toronto, Canada.* 16. Toronto, Canada; 2008:2972.

[34] Li Y, Xie Z, Pang Y, Vigneron D, Zhang X. ICE decoupling technique for RF coil array designs. Med Phys 2011;38(7):4086-93.

[35] Yan X, Cao Z, Zhang X. Simulation verification of SNR and parallel imaging improvements by ICE-decoupled loop array in MRI. Appl Magn Reson 2016;47(4):395-403.

[36] Li Y, Pang Y, Vigneron D, Glenn O, Xu D, Zhang X. Investigation of multichannel phased array performance for fetal MR imaging on 1.5T clinical MR system. Quant Imaging Med Surg 2011;1:24-30.

[37] Yan X, Wei L, Chu S, Xue R, Zhang X. Eight-Channel Monopole Array Using ICE Decoupling for Human Head MR Imaging at 7 T. Appl Magn Reson 2016;47(5):527-38.

[38] Yan X, Xie Z, Pedersen JO, Zhang X. Theoretical analysis of magnetic wall decoupling method for radiative antenna arrays in ultrahigh magnetic field MRI. Concepts in Magnetic Resonance Part B 2015;45B(4):183–90.

[39] Payne K, Khan SI, Zhang X. Investigation of Magnetic Wall Decoupling for planar Quadrature RF Array coils using Common-Mode Differential-mode Resonators. Proc Int Soc Magn Reson Med Sci Meet Exhib 2022;30.

[40] Payne K, Bhosale AA, Zhang X. Double cross magnetic wall decoupling for quadrature transceiver RF array coils using common-mode differential-mode resonators. J Magn Reson 2023;353:107498.

[41] Payne K, Bhosale AA, Ying LL, Zhang X. Quadrature Transceiver RF Arrays Using Double Cross Magnetic Wall Decoupling for Ultrahigh field MR Imaging. Proc Int Soc Magn Reson Med Sci Meet Exhib Int Soc Magn Reson Med Sci Meet Exhib 2023;31.

[42] Yan X, Gore JC, Grissom WA. Self-decoupled radiofrequency coils for magnetic resonance imaging. Nat Commun 2018;9(1):3481.

[43] Zhang X, DelaBarre L, Payne K, Waks M, Adriany G, Ugurbil K. A Wrap-on Decoupled Coaxial Transmission Line (CTL) Transceive Array for Head MR Imaging at 10.5T. *Proc Intl Soc Mag Reson Med.* 2023:3904.

[44] Zhang X, Waks M, DelaBarre L, Payne K, Ugurbil K, Adriany G. Design and Test of a Flexible Two-row CTL Array and Its Detunable Resonant Elements for 10.5T MR Imaging. *Proc Intl Soc Mag Reson Med.* 2023:4593.

[45] Payne K, Ying LL, Zhang X. Hairpin RF resonators for MR imaging transceiver arrays with high inter-channel isolation and B1 efficiency at ultrahigh field 7 T Journal of Magnetic Resonance 2022;345:107321.

[46] Payne K, Zhang X. Quadrature RF array using High Impedance concept for improved transmit RF field B1 efficiencyat 7 Tesla. Proc Int Soc Magn Reson Med Sci Meet Exhib Int Soc Magn Reson Med Sci Meet Exhib 2022;30.

[47] Collins CM, Wang Z. Calculation of radiofrequency electromagnetic fields and their effects in MRI of human subjects. Magn Reson Med 2011;65(5):1470-82.

[48] Knecht O, Kolar JW. Impact of Transcutaneous Energy Transfer on the Electric Field and Specific Absorption Rate in the Human Tissue. *the 41th Annual Conference of the IEEE Industrial Electronics Society (IECON 2015).* Yokohama, Japan: IEEE; 2015.

[49] Teeuwisse WM, Brink WM, Haines KN, Webb AG. Simulations of high permittivity materials for 7 T

neuroimaging and evaluation of a new barium titanate-based dielectric. Magn Reson Med 2012;67(4):912-8.

[50] Haines K, Smith NB, Webb AG. New high dielectric constant materials for tailoring the B1+ distribution at high magnetic fields. J Magn Reson 2010;203(2):323-7.

[51] Snaar JE, Teeuwisse WM, Versluis MJ, van Buchem MA, Kan HE, Smith NB, et al. Improvements in high-field localized MRS of the medial temporal lobe in humans using new deformable high-dielectric materials. NMR Biomed 2011;24(7):873-9.

[52] Wang C, Zhang X. Evaluation of B1+ and E field of RF Resonator with High Dielectric Insert. *Proceedings of the 17th Annual Meeting of ISMRM, Honolulu, Hawaii.* 2009:3054.

[53] Yang QX, Mao W, Wang J, Smith MB, Lei H, Zhang X, et al. Manipulation of image intensity distribution at 7.0 T: passive RF shimming and focusing with dielectric materials. J Magn Reson Imaging 2006;24(1):197-202.

[54] Yang QX, Wang J, Zhang X, Collins CM, Smith MB, Liu H, et al. Analysis of wave behavior in lossy dielectric samples at high field. Magn Reson Med 2002;47(5):982-9.

[55] Bhosale A, Zhang X. High dielectric sheet to reduce electric fields in Self-decoupled radiofrequency coils for magnetic resonance imaging. *Proceedings of the 29th Annual Meeting of ISMRM.* 2021.

[56] Bhosale AA, Gawande D, Zhang X. B1 field flattening and length control of half-wave dipole antenna with discrete dielectric coating. Proc Int Soc Magn Reson Med Sci Meet Exhib Int Soc Magn Reson Med Sci Meet Exhib 2022;30.

[57] Bhosale AA, Gawande D, Zhang X. A Dielectric Material Coated Half-Wave Dipole antenna for Ultrahigh Field MRI at 7T/300MHz. Proc Int Soc Magn Reson Med Sci Meet Exhib Int Soc Magn Reson Med Sci Meet Exhib 2022;30.

[58] Bhosale AA, Ying LL, Zhang X. An 8-Channel High-permittivity Dielectric Material-Coated Half-Wave Dipole Antenna Array for Knee Imaging at 7T. Proc Int Soc Magn Reson Med Sci Meet Exhib Int Soc Magn Reson Med Sci Meet Exhib 2022;30.

[59] Bhosale AA, Ying LL, Zhang X. Design of a 13-Channel hybrid RF array with field rectification of dielectric material for foot/ankle imaging at 7T. Proc Int Soc Magn Reson Med Sci Meet Exhib Int Soc Magn Reson Med Sci Meet Exhib 2022;30.